\documentclass[12pt,preprint]{aastex}

\newcommand{\msun} {$M_\odot$}
\shorttitle{Companion to the Lowest Mass White Dwarf}
\shortauthors{Kilic et al.}

\begin{document}

\title{The Discovery of a Companion to the Lowest Mass White Dwarf\footnote{Observations reported here were obtained at the MMT Observatory, a joint facility of the Smithsonian Institution and the University of Arizona.}}

\author{Mukremin Kilic\altaffilmark{2}, Warren R. Brown\altaffilmark{3}, Carlos Allende Prieto\altaffilmark{4}, M. H. Pinsonneault\altaffilmark{2}, and S. J. Kenyon\altaffilmark{3}}

\altaffiltext{2}{Ohio State University, Department of Astronomy, 140 West 18th Avenue, Columbus, OH 43210}

\altaffiltext{3}{Smithsonian Astrophysical Observatory, 60 Garden Street, Cambridge, MA 02138}

\altaffiltext{4}{McDonald Observatory and Department of Astronomy, University of Texas, Austin, TX 78712}

\begin{abstract}

We report the detection of a radial velocity companion to SDSS J091709.55+463821.8, the lowest mass white dwarf currently known with $M\sim 0.17$\msun.
The radial velocity of the white dwarf shows variations with a semi-amplitude of 148.8 $\pm$ 6.9 km s$^{-1}$ and
a period of 7.5936 $\pm$ 0.0024 hours, which implies a companion mass of $M\geq0.28$\msun. 
The lack of evidence of a companion in the optical photometry forces any main-sequence companion to be smaller than 0.1\msun, hence
a low mass main sequence star companion is ruled out for this system.
The companion is most likely another white dwarf, and we present tentative evidence for an evolutionary scenario which could have produced it.
However, a neutron star companion cannot be ruled out and follow-up radio observations are required to search for a pulsar companion. 

\end{abstract}

\keywords{stars: individual (SDSS J091709.55+463821.8) -- low-mass -- white dwarfs}

\section{Introduction}

Recent discoveries of several extremely low mass white dwarfs (WDs) in the field (Kilic et al. 2007; Eisenstein et al. 2006; Kawka et al. 2006)
and around pulsars (Bassa et al. 2006; van Kerkwijk et al. 1996) show that WDs with mass as low as 0.17\msun\ are formed in
the Galaxy. No galaxy is old enough to produce such extremely low mass WDs through single star evolution. The oldest globular clusters in our
Galaxy are currently producing $\sim0.5$\msun\ WDs (Hansen et al. 2007), therefore lower mass WDs must experience significant
mass loss.  The most likely explanation is a close binary companion. If a WD forms in a close binary, it can lose its outer envelope without reaching
the asymptotic giant branch and without ever igniting helium, ending up as a low mass, helium core WD. Confirmation of the binary nature
of several low mass WDs by Marsh et al. (1995) supports this binary formation scenario. 

White dwarf binaries provide an important tool for testing binary evolution, specifically the efficiency of the mass loss process
and the common envelope phase. Since WDs can be created only at the cores of giant stars, their properties can be used to reconstruct
the properties of the progenitor binary systems. Using a simple core mass - radius relation for giants and the known orbital period,
the initial orbital parameters of the binary system can be determined. 
For a review of binary evolution involving WDs, see e.g. Sarna et al. (1996),
Iben et al. (1997), Sandquist et al. (2000), Yungelson et al. (2000), Nelemans \& Tout (2005), and Benvenuto \& De Vito (2005).

Known companions to low mass WDs include late type main sequence stars (Farihi et al. 2005; Maxted et al. 2007), helium or carbon/oxygen core WDs
(Marsh et al. 1995; Marsh 2000; Napiwotzki et al. 2001), and in some cases neutron stars (Nice et al. 2005).
Late type stellar companions to low mass WDs have a distribution of masses with median 0.27\msun\ (Nelemans \& Tout 2005).
This median companion mass is nearly identical to the peak companion mass of 0.3\msun (spectral type M3.5) observed in the field population 
of late type main sequence stars within 20 pc of the Earth (Farihi et al. 2005).  
Low mass WD - WD binaries, on the other hand, tend to have equal mass WDs.
The median mass for both the brighter and dimmer components of the known low mass WD binary systems is 0.44\msun\ (Nelemans \& Tout 2005).

The discovery of extremely low mass WDs around pulsars suggests that neutron star companions may be responsible for
creating white dwarfs with masses of about 0.2\msun. PSR J0437-4715, J0751+1807, J1012+5307, J1713+0747, B1855+09, and J1909-3744 are pulsars in
pulsar -- He-WD binary systems with circular orbits and orbital periods of $\sim$0.2-100 days (Nice et al. 2005).
So far, only two of these companions are spectroscopically confirmed to be helium-core WDs.
Van Leeuwen et al. (2007) searched for radio pulsars around 8 low mass WDs, but did not find any companions. They
concluded that the fraction of low mass helium-core WDs with neutron star companions is less than 18\% $\pm$ 5\%.

Kilic et al. (2007) reported the discovery of the lowest mass WD currently known: SDSS J091709.55+463821.8 (hereafter J0917+46).
With an estimated mass of $0.17$\msun, J0917+46 provides a unique opportunity to search for a binary companion and to test our understanding
of the formation scenarios for extremely low mass WDs. Do extremely low mass WDs form in binaries with neutron stars, WDs, or late type stars?
If the companion is a neutron star, the mass of the neutron star can be used to constrain the neutron star equation of state.
In case of a WD or a late type star companion, the orbital parameters can be used to constrain the common-envelope phase of binary
evolution in these systems. In this Letter, we present new optical spectroscopy and radial velocity measurements for J0917+46.
Our observations are discussed in \S 2, while an analysis of the spectroscopic data and the discovery of a companion are discussed in \S 3.
The nature of the companion is discussed in \S 4.

\section{Observations}

We used the 6.5m MMT telescope equipped with the Blue Channel Spectrograph to obtain moderate
resolution spectroscopy of SDSS J0917+46 nine times over the course of five nights between UT 2006 December 22-27 and five times on UT 2007 March 19.
The spectrograph was operated with the 832 line mm$^{-1}$ grating in second order, providing a wavelength
coverage of 3650 - 4500 \AA. Most spectra were obtained with a 1.0$\arcsec$ slit yielding a resolving power of $R=$ 4300,
however a 1.25$\arcsec$ slit was used on 2006 December 24, which resulted in a resolving power of 3500.
Exposure times ranged from 15 to 22 minutes and yielded signal-to-noise ratio $S/N > 20$ in the continuum at 4000 \AA.
All spectra were obtained at the parallactic angle, and comparison lamp exposures were obtained after every exposure.
The spectra were flux-calibrated using blue spectrophotometric standards (Massey et al. 1988). 

Heliocentric radial velocities were measured using the cross-correlation package RVSAO (Kurtz \& Mink 1998).
We obtained preliminary velocities by cross-correlating the observations with bright WD templates of known velocity.
However, greater velocity precision comes from cross-correlating J0917+46 with itself.
Thus we shifted the individual spectra to rest-frame and summed them together into a high S/N template spectrum.
Our final velocities come from cross-correlating the individual observations with the J0917+46 template, and are presented in Table 1.

\section{The Discovery of a Companion}

The radial velocity of the lowest mass WD varies by as much as 263 km s$^{-1}$ between different observations, 
revealing the presence of a companion object.
We solve for the best-fit orbit using the code of Kenyon \& Garcia (1986).
We find that the heliocentric radial velocities of the WD are best fitted with a circular orbit and a radial velocity amplitude
K = 148.8 $\pm$ 6.9 km s$^{-1}$. We used the method of Lucy \& Sweeney (1971) to show that there is no evidence for an eccentric orbit
from our data and that the 1$\sigma$ upper limit to the eccentricity is $e=0.06$. We have assumed that the orbit is circular.
The orbital period is 7.5936 $\pm$ 0.0024 hours with spectroscopic conjunction at HJD 2454091.73 $\pm$ 0.002.
Figure 1 shows the observed radial velocities and our best fit model for SDSS J0917+46.

Measurement of the orbital period $P$ and the semi-amplitude of the radial velocity variations $K$ allows us to calculate the mass function

\begin{equation} 
\frac{M^3~{\rm sin}^3i}{(M_{\rm WD}+M)^2}=\frac{P K^3}{2 \pi G}= 0.108 \pm 0.018 M_\sun,
\end{equation} 

\noindent where $i$ is the orbital inclination angle, $M_{\rm WD}$ is the WD mass (0.17\msun\ for SDSS J0917+46), and $M$ is the mass of the companion object.
We can put a lower limit on the mass of the companion by assuming an edge-on orbit (sin $i=$ 1), for which the companion would be
an $M=0.28$\msun\ object at an orbital separation of 1.5$R_{\sun}$. Therefore, the companion mass is $M\geq0.28$\msun.

\section{The Nature of the Companion}

To understand the nature of the companion, first we need to understand the properties of the WD. Kilic et al.'s (2007) analysis
of J0917+46 was based on a single 15 minute exposure with S/N = 20 at 4000 \AA. Here we use 14 different spectra of J0917+46 (4.3 hours total exposure time) 
and repeat our spectroscopic analysis. We also combine all of these spectra into a composite (weighted average) spectrum with $R=3500$ that results in S/N = 80 at 4000 \AA.
Figure 2 shows the composite spectrum and our fits using the entire spectrum (excluding the Ca K line) and also using only the Balmer lines (see Kilic et al. 2007 for the details of the spectroscopic analysis).
We find a best-fit solution of $T_{\rm eff}=11855$ K and $\log$ g = 5.55 if we use the observed composite spectrum. If we normalize (continuum-correct) the
composite spectrum and fit just the Balmer lines, then we obtain $T_{\rm eff} =11817$ K and $\log$ g = 5.51. Since we have 14 different spectra, we also fit each
spectrum individually to obtain a robust estimate of the errors in our analysis. For the first case where we fit the observed spectra, we obtain a best fit solution of
$T_{\rm eff} = 11984 \pm 168$ K and $\log$ g = 5.57 $\pm 0.05$. For the second case where we fit
only the Balmer lines, we obtain $T_{\rm eff} = 11811 \pm 67$ K and $\log$ g = 5.51 $\pm 0.02$. 
Our results are consistent with each other and also with Kilic et al.'s analysis. We confirm that our temperature and gravity estimates are robust;
SDSS J0917+46 is still the lowest gravity/mass WD currently known. 

We adopt our best fit solution of $T_{\rm eff} = 11855 \pm 168$ K and $\log$ g = 5.55 $\pm 0.05$ for the remainder of the paper.
Using our new temperature and gravity measurements and Althaus et al. (2001) models, we estimate the absolute magnitude of the
WD to be $M_{\rm V}\sim$ 7.0, corresponding to a distance modulus of 11.8 (at 2.3 kpc) and a cooling age of about 500 Myr.
At a Galactic latitude of $+44^{\circ}$, J0917+46 is located at 1.6 kpc above the plane. The radial velocity of the binary system is $\sim$29 km s$^{-1}$.
J0917+46 displays a proper motion of $\mu_{\rm RA} cos \delta = -2 \pm 3.4$ mas yr$^{-1}$ and $\mu_{\rm DEC}= 2 \pm 3.4$ mas yr$^{-1}$ measured from its SDSS and USNO-B positions
(kindly provided by J. Munn). These measurements correspond to a tangential velocity of 31 $\pm$ 52 km s$^{-1}$.
J0917+46 has disk kinematics, and its location above the Galactic plane is consistent with a thick disk origin.
Therefore the main sequence age of the progenitor star of the lowest mass WD needs to be on the order of 10 Gyr; the progenitor was a $\sim$1\msun\ main sequence star.

The broad-band spectral energy distribution of J0917+46 is shown in panel c in Figure 2. The de-reddened SDSS photometry and the fluxes
predicted for the parameters derived from our spectroscopic analysis are shown as error bars and circles, respectively.
The SDSS photometry is consistent with our spectroscopic solution within the errors. The $g$-band photometry is slightly brighter
than expected, however it is within 1.7$\sigma$ of our spectroscopic solution and therefore the excess is not significant. We also note that
many low mass WD candidates analyzed by Kilic et al. (2007) had discrepant $g$-band photometry. A similar problem may cause the
observed slight $g$-band excess.
 
The mass function for our target plus the MMT spectroscopy and the SDSS photometry can be used to constrain
the nature of the companion star. Since the companion mass has to be $\geq0.28$\msun, it can be a low mass star, another WD, or a neutron star.

\subsection{A Low Mass Star}

\subsubsection{Constraints from the SDSS Photometry}

If the orbital inclination angle of the binary system is between 47$^{\circ}$ and 90$^{\circ}$, the companion mass would be 0.28-0.50\msun, consistent
with being an M dwarf. However, an M dwarf companion would show up as an excess in the SDSS photometry. For example, an M6 dwarf would cause a 10\% excess
in the z-band photometry of J0917+46 at its distance, and hence any star with $M\geq0.1$\msun\ would be visible in the SDSS photometry. Since the mass function
derived from the observed orbital parameters of the binary limits the companion mass to $\geq0.28$\msun\ (earlier than M3.5V spectral type), and any such
companion would be visible in the SDSS photometry (see panel c in Figure 2), a main sequence star companion is ruled out.

\subsubsection{Constraints from the MMT Spectroscopy}

We search for spectral features from a companion by subtracting all 14 individual spectra from our best-fit WD model after shifting the
individual spectra to the rest-frame. The {\it only} significant feature that we detect is a Ca K line. This same calcium line was also present in the discovery spectrum
of J0917+46 (Kilic et al. 2007). It had the same radial velocity as the Balmer lines and hence, it was predicted to be photospheric. The
Ca K line is visible in all of our new spectra of this object, and its radial velocity changes with the radial velocity of the Balmer lines;
it is confirmed to be photospheric. The Ca H line overlaps with the saturated H$\epsilon$ line and it is not detected in our spectrum.
If J0917+46 had a low mass star companion, it could contribute an additional calcium line in the spectrum. 

Marsh et al. (1995) were able to find the companion to the low mass white dwarf WD1713+332 by searching for asymmetries in the H$\alpha$ line profile. We perform a similar
analysis for J0917+46 using the calcium line.
We combine the spectra near maximum radial velocity ($V\geq125$ km s$^{-1}$) and near minimum radial velocity ($V\leq-80$ km s$^{-1}$) into two composite spectra. If there
is a contribution from a companion object, it should be visible as an asymmetry in the line profile. This asymmetry should be on the blue side of the red-shifted composite spectrum,
and on the red side of the blue-shifted composite spectrum. Figure 3 shows the red-shifted (solid line) and blue-shifted (dotted line) composite spectra of J0917+46. The rest-frame wavelegth of the Ca K line is
shown as a dashed line. This figure demonstrates that there is an asymmetry in the Ca K line profile of the red-shifted spectrum, however the additional Ca K feature is close to
the rest-frame velocity, and it is not detected in the blue-shifted spectrum. The equivalent width of the main Ca K feature in the red-shifted spectrum is 0.42\AA, and the additional
feature has an equivalent width of $\sim$0.16 \AA. The blue shifted spectrum has a stronger Ca K feature with an equivalent width of 0.57 \AA, and it is consistent with a blend of
the two Ca K features seen in the red-shifted spectrum. The observed additional calcium feature seems to be stationary and it seems to originate from
interstellar absorption. Therefore, we conclude that our optical spectroscopy does not reveal any spectral features from a companion object.

\subsection{Another White Dwarf}

Using the most likely inclination angle for a random stellar sample, $i=60^{\circ}$, we estimate that the companion mass is most likely to be 0.36\msun, another low mass WD.
The orbital separation of the system would be about 1.6$R_{\sun}$. If the inclination angle is between 47$^{\circ}$ and 27$^{\circ}$ (21\% likelihood), the companion mass would be
0.5-1.4\msun, consistent with a normal carbon/oxygen or a massive WD.

Liebert et al. (2004) argue that it is possible to create a 0.16-0.19\msun\ WD from a 1\msun\ progenitor star, if the binary separation is appropriate and the common-envelope phase is sufficiently unstable so that the envelope can be lost quickly from the system.  
We re-visit their claim to see if J0917+46 can be a low mass WD formed from such a progenitor system.

Close WD pairs can be created by two consecutive common envelope phases or an Algol-like stable mass transfer phase followed by a common envelope phase (Iben et al. 1997).
In the first scenario, due to orbital shrinkage the recently formed WD is expected to be less massive than its companion by a factor of $\leq$0.55.
On the other hand, the expected mass ratio for the second scenario involving a stable mass transfer and a common-envelope phase is around 1.1 (Nelemans et al. 2000).
The mass ratio of the J0917+46 binary is $M_{\rm bright}/M_{\rm dim}\leq0.61$, therefore the progenitor binary star system has probably gone through two common envelope phases.

Nelemans \& Tout (2005) found that the common-envelope evolution of close WD binaries can be reconstructed with an algorithm ($\gamma$-algorithm) equating
the angular momentum balance. The final orbital separation of a binary system that went through a common envelope phase is

\begin{equation}
\frac{a_{\rm final}}{a_{\rm initial}} = \left(\frac{M_{\rm giant}}{M_{\rm core}}\right)^2 \left(\frac{M_{\rm core} + M_{\rm companion}}{M_{\rm giant} + M_{\rm companion}}\right)  \left(1 - \gamma \frac{M_{\rm envelope}}{M_{\rm giant} + M_{\rm companion}}\right)^2,
\end{equation}

\noindent where a is the orbital separation, $M$ are the masses of the companion, core, giant, and the envelope, and $\gamma=1.5$ (Nelemans \& Tout 2005).

We assume that the mass of the WD is the same as the mass of the core of the giant at the onset of the mass transfer. 
Assuming a giant mass of 0.8-1.3\msun, a core mass of 0.17\msun, and possible WD companion masses of 0.28-1.39\msun,
we estimate the initial orbital separation.
Using the core-mass - radius relation for giants, $R = 10^{3.5} M_{\rm core}^4$ (Iben \& Tutukov 1985), we find that the radius
of the giant star that created J0917+46 was R = $2.6R_{\sun}$. For $M$ = 0.28-1.39\msun\ companions, we use the size
of the Roche lobe $R_{\rm L}$ as given by Eggleton (1983) to determine the separation at the onset of mass transfer assuming $R_{\rm giant} = R_{\rm L}$.
The size of the Roche lobe depends on the mass ratio and the orbital separation of the system.
The initial separation and the size of the Roche lobe gives us a unique solution for the binary mass ratio.
Table 2 presents the companion mass, initial orbital separation and the orbital period for 0.8-1.3\msun\ giants that could
create the lowest mass WD with the observed orbital parameters.

The mass function for J0917+46 favors a low mass WD companion, therefore the first four scenarios in Table 2 seem to be more likely.
For example, a 0.8\msun\ giant and a 0.33\msun\ WD companion with an initial orbital separation of 5.7$R_{\sun}$ and an orbital period of 36 hr would create
a 0.17\msun\ WD with the observed orbital parameters.

The same procedures can be used to re-create the first common-envelope phase of the binary evolution. The progenitor of the unseen companion to J0917+46
must be in the range 0.8-2.3\msun. The lower mass limit is set by the fact that the unseen companion has to be more massive than the lowest mass WD. The upper limit
is set by the fact that more massive stars do not form degenerate helium cores (Nelemans et al. 2000) and that a common envelope phase with a more massive giant would
end up in a merger and not in a binary system.
Assuming 0.8-2.3\msun\ giants and 0.8-1.3\msun\ main sequence companions, we calculate possible evolutionary scenarios to create the orbital parameters of the
binary system before the last common envelope phase. We find that a 2.2\msun\ star and a 0.8\msun\ companion (which is the progenitor of the lowest mass WD)
with an orbital period of 51 days and orbital separation of 0.4 AU would create a 0.33\msun\ WD. In addition, we find that the progenitor of the lowest mass WD 
has to be less massive than 0.9\msun\, since a $M\geq$ 0.9\msun\ progenitor would require a companion more massive than 2.3\msun\ to match the binary properties of the system before the
last common envelope phase. Therefore, the most likely evolutionary scenario for a WD + WD binary involving
J0917+46 would be: 2.2\msun\ giant + 0.8\msun\ star $\longrightarrow$ 0.33\msun\ WD + 0.8\msun\ star $\longrightarrow$ 0.33\msun\ WD
+ 0.8\msun\ giant $\longrightarrow$ 0.33\msun\ WD + 0.17\msun\ WD. 

The main sequence lifetime of a 2.2\msun\ thick disk ([Fe/H] $\simeq -0.7$) star is about 650 Myr (Bertelli et al. 1994), and a 0.33\msun\ white
dwarf created from such a system would be a $\sim$10 Gyr old WD. According to Althaus et al. (2001) models, a 0.33\msun\ He-core WD would cool down to
3700 K in 10 Gyr and it would have $M_V\sim15.8$. This companion would be several orders of magnitude fainter than the 0.17\msun\ WD observed today, and therefore
the lack of evidence of a companion in the SDSS photometry and our optical spectroscopy is consistent with this formation scenario.

\subsection{A Neutron Star}

The formation scenarios for low mass helium WDs in close binary systems also include neutron star companions. There are already several low mass
WDs known to have neutron star companions (van Kerkwijk et al. 1996; Nice et al. 2005; Bassa et al. 2006). According to the theoretical calculations
by Benvenuto \& De Vito (2005), a 1\msun\ star with a neutron star companion in an initial orbital period of 0.5-0.75 days would end up as a
0.05-0.21\msun\ WD with a 1.8-1.9\msun\ neutron star in an orbital period of 0.03-3.5 days. They expect a 0.17\msun\ WD + neutron star binary
to have an orbital period of 9.6 hours (see their Figure 16). The orbital period of J0917+46 is consistent with their analysis.

If the orbital inclination angle of the J0917+46 binary system is less than 27$^{\circ}$, the companion mass would be $\geq$1.4\msun, consistent with being a neutron star.
The probability of observing a binary system at an angle less than 27$^{\circ}$ is only 11\%. This is unlikely, but cannot be definitely ruled out. 

\section{Discussion}

Our radial velocity measurements of J0917+46 shows that it is in a binary system with an orbital period of 7.59 hours.
Short period binaries may merge within a Hubble time by losing angular momentum through gravitational radiation. The merger time for such binaries is given by
\begin{equation}
\tau = \frac{(M_1 + M_2)^{1/3}}{M_1 M_2} P^{8/3} \times 10^7 yr
\end{equation}

\noindent where the masses are in solar units and the period is in hours (Landau \& Lifshitz 1958; Marsh et al. 1995). For the J0917+46 binary, if the companion is a low mass
WD with $M\leq0.5$\msun, the merger time is longer than 23 Gyr. If the companion is a 1.4\msun\ neutron star, then the merger time would be
10.8 Gyr. J0917+46 binary system will not merge within the next 10 Gyr.

J0917+46 has nearly solar calcium abundance, and it has more calcium than many of the metal-rich WDs with circumstellar debris disks.
The extremely short timescales for gravitational settling of Ca implies that the WD requires an external source for the observed metals (Koester \& Wilken 2006;
Kilic \& Redfield 2007). The star is located far above the Galactic plane where accretion from the ISM
is unlikely. A possible scenario for explaining the photospheric calcium is accretion from a circumbinary disk created during the mass loss phase of the giant.
Since J0917+46 went through a common envelope phase with a companion, a left-over circumbinary disk is possible.
An accretion disk is observed around the companion to the giant Mira A (Ireland et al. 2007). 
In addition, a fallback disk is observed around a young neutron star (Wang et al. 2006). A similar mechanism could create a circumbinary disk around J0917+46. 

Several calcium-rich WDs are known to host disks (Kilic et al. 2006; von Hippel et al. 2007). These WDs are 0.2 - 1.0 Gyr old. The presumed origin of these disks
is tidal disruption of asteroids or comets, however a leftover disk from the giant phase of the WD is not completely ruled out (von Hippel et al. 2007).
Su et al. (2007) detected a disk around the central star of the Helix planetary nebula. This disk could
form from the left-over material from the giant phase that is ejected with less than the escape speed. 
The majority of the metal-rich white dwarfs with disks show 30-100\% excess in the K-band (Kilic \& Redfield 2007). J0917+46 is expected to be fainter than
19th magnitude in the near-infrared, and it is not detected in the Point Source Catalog (PSC) part of
the Two Micron All Sky Survey (2MASS;  Skrutskie et al. 2006). The PSC 99\% completeness limits are 15.8, 15.1 and 14.3 in $J, H$
and $K_s$ filters, respectively. These limits do not provide any additional constraints on the existence of a disk or a companion object.
Follow-up near-infrared observations with an 8m class telescope are required to search for the signature of a debris disk which could explain
the observed calcium abundance in this WD.

\section{Conclusions}

SDSS J0917+46, the lowest gravity/mass WD currently known, has a radial velocity companion.
The lack of excess in the SDSS photometry and the orbital parameters of the system rule out a low mass star companion. We find that the companion is likely to be
another WD, and most likely to be a low mass WD. We show that if the binary separation is appropriate and the common-envelope
phase is efficient, it is possible to create a 0.17\msun\ WD in a 7.59 hr orbit around another WD. A neutron star companion
is also possible if the inclination angle is smaller than 27$^\circ$; the likelihood of this is 11\%.
If the companion is a neutron star, it would be a milli-second pulsar. Radio observations of J0917+46 are needed to search for such a companion.

\acknowledgements
M. Kilic thanks A. Gould for helpful discussions.

{\it Facilities:} \facility{MMT (Blue Channel Spectrograph)}

\clearpage
\begin{deluxetable}{cr}
\tablecolumns{2}
\tablewidth{0pt}
\tablecaption{Radial Velocity Measurements for SDSS J0917+46}
\tablehead{
\colhead{Julian Date}&
\colhead{Heliocentric Radial Velocity}\\
 & (km s$^{-1}$)
}
\startdata
2454091.77060 &   134.29 $\pm$ 5.45 \\
2454091.85741 &   124.83 $\pm$ 8.56 \\
2454093.82302 & $-$82.91 $\pm$ 4.22 \\
2454093.94566 &    41.98 $\pm$ 4.65 \\
2454094.04293 &   152.42 $\pm$ 3.79 \\
2454095.83998 &    39.88 $\pm$ 3.45 \\
2454095.89902 &   172.60 $\pm$ 3.57 \\
2454096.04373 & $-$79.87 $\pm$ 3.76 \\
2454096.95394 &    20.45 $\pm$ 7.49\\
2454178.62439 & $-$86.78 $\pm$ 6.08\\
2454178.69885 & $-$90.08 $\pm$ 3.90\\
2454178.72963 & $-$11.16 $\pm$ 6.32\\
2454178.77125 &   106.28 $\pm$ 6.53\\
2454178.83047 &   160.16 $\pm$ 5.48
\enddata
\end{deluxetable}

\begin{deluxetable}{ccccccc}
\tablecolumns{7}
\tablewidth{0pt}
\tablecaption{The Last and the First Common-Envelope Phases}
\tablehead{
\colhead{CE Phase}&
\colhead{M$_{\rm giant}$}&
\colhead{M$_{\rm companion}$}&
\colhead{a$_{\rm initial}$}&
\colhead{a$_{\rm final}$}&
\colhead{P$_{\rm initial}$}&
\colhead{P$_{\rm final}$}\\
& (\msun) & (\msun) & ($R_{\sun}$) & ($R_{\sun}$) & (hr) & (hr)
}
\startdata
2 & 0.8 &  0.33 & 5.70 & 1.55 & 35.6 & 7.6 \\
2 & 0.9 &  0.39 & 5.73 & 1.61 & 33.6 & 7.6 \\
2 & 1.0 &  0.45 & 5.78 & 1.66 & 32.1 & 7.6 \\
2 & 1.1 &  0.50 & 5.78 & 1.71 & 30.5 & 7.6 \\
2 & 1.2 &  0.56 & 5.85 & 1.75 & 29.7 & 7.6 \\
2 & 1.3 &  0.61 & 5.84 & 1.80 & 28.4 & 7.6 \\
\hline
1 & 2.2 &  0.80 & 83.52 & 5.70 & 1222.9 & 35.6
\enddata
\end{deluxetable}

\clearpage
\begin{figure}
\plotone{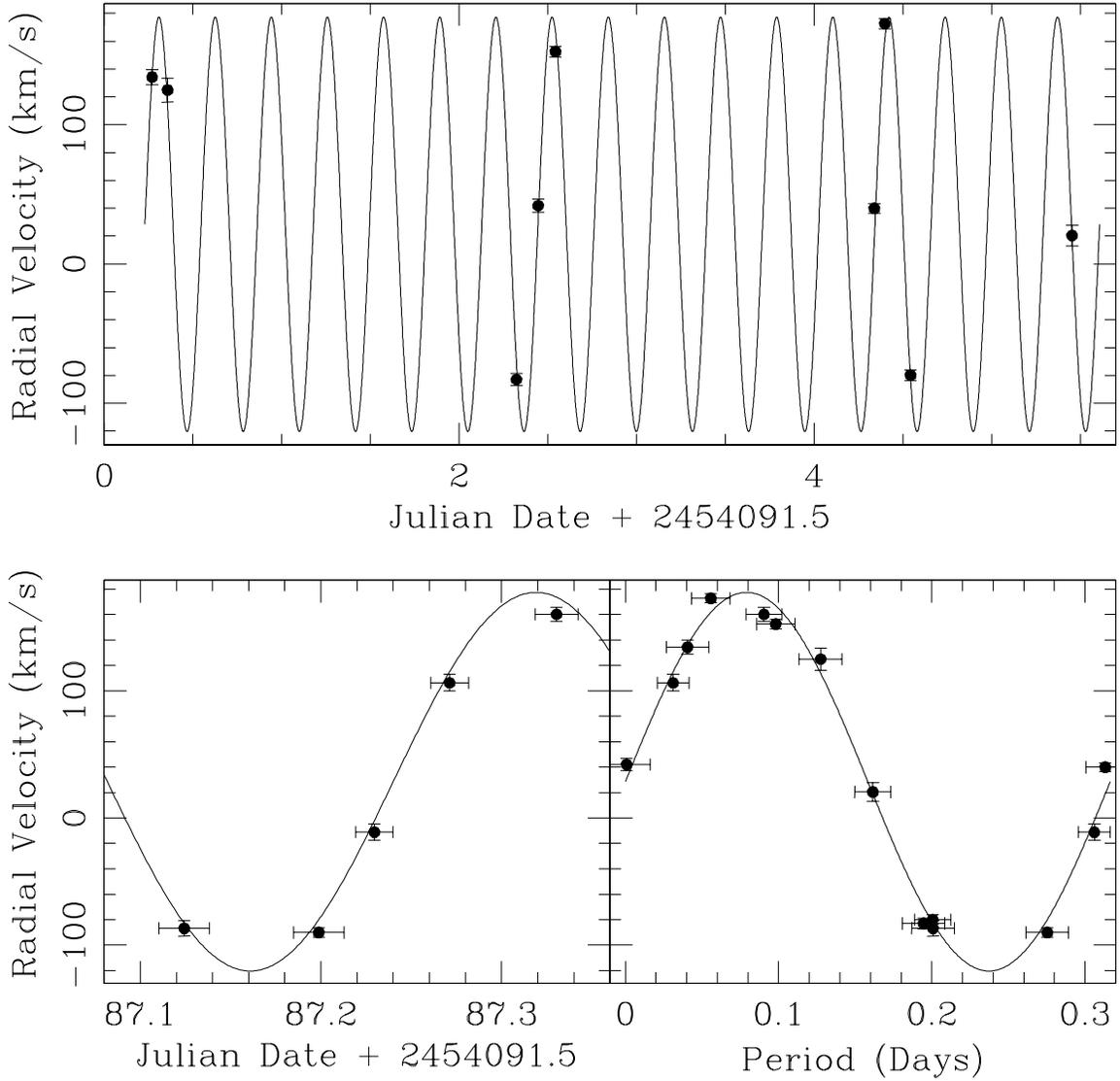}                
\caption{The radial velocities of the white dwarf SDSS J0917+46 (black dots) observed in 2006 December (top panel) and 2007 March (bottom left panel).
The bottom right panel shows all of these data points phased with the best-fit period. The solid line represents the best-fit model for a circular orbit with a radial
velocity amplitude of 148.8 km s$^{-1}$ and a period of 7.5936 hours.}
\end{figure}

\clearpage
\begin{figure}
\includegraphics[angle=-90,scale=.7]{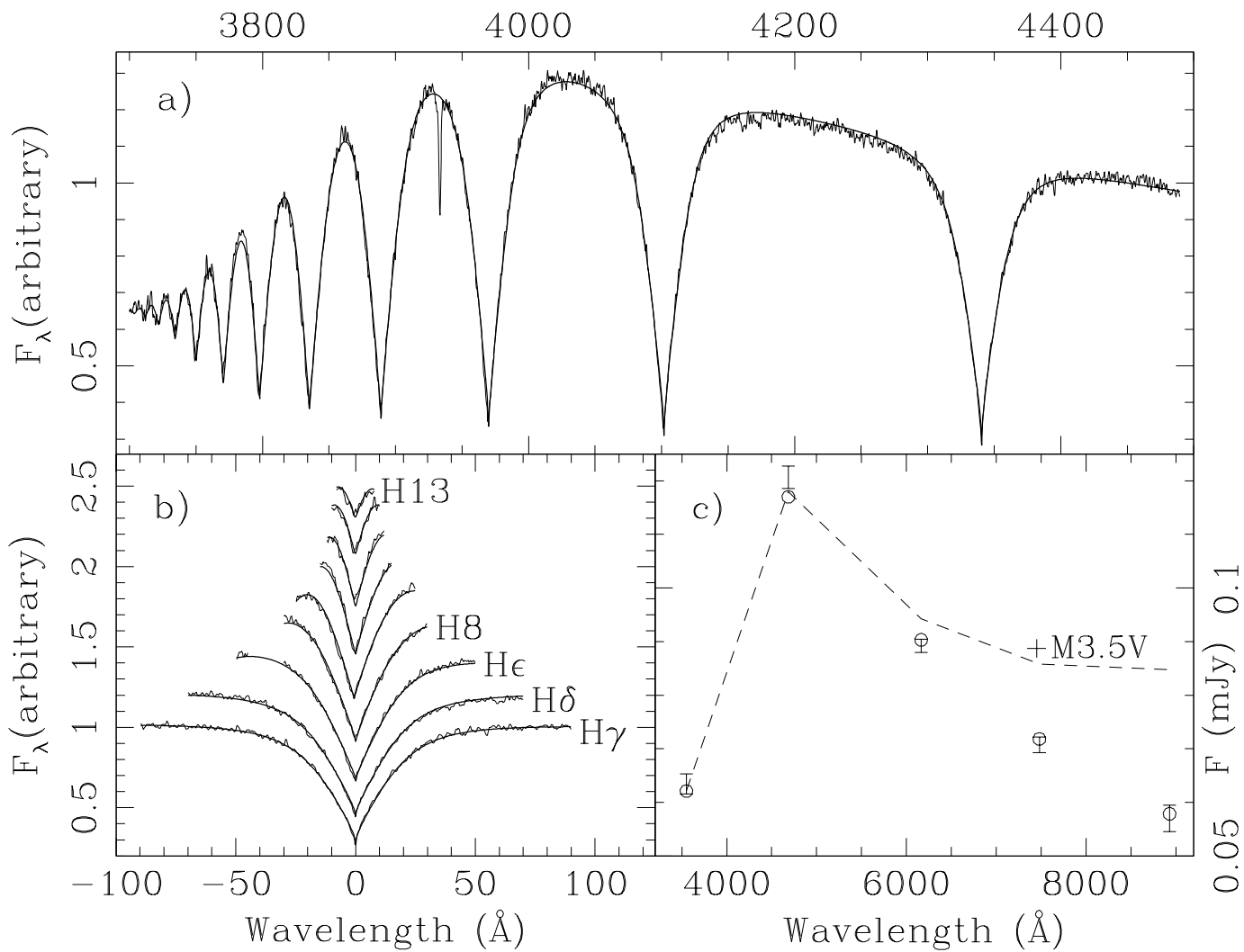} 
\caption{Spectral fits (solid lines) to the observed composite spectrum of SDSS J0917+46 (jagged lines, panel a) and to the flux-normalized line profiles
(panel b). The Ca K line region (3925 - 3940 \AA) is not included in our fits. The SDSS photometry (error bars) and the predicted fluxes from our best fit solution to the spectra (circles) are shown in panel c. The dashed line shows the effect of adding an M3.5V (0.3\msun) companion to our best-fit white dwarf model.}
\end{figure}

\clearpage
\begin{figure}
\includegraphics[angle=-90,scale=.65]{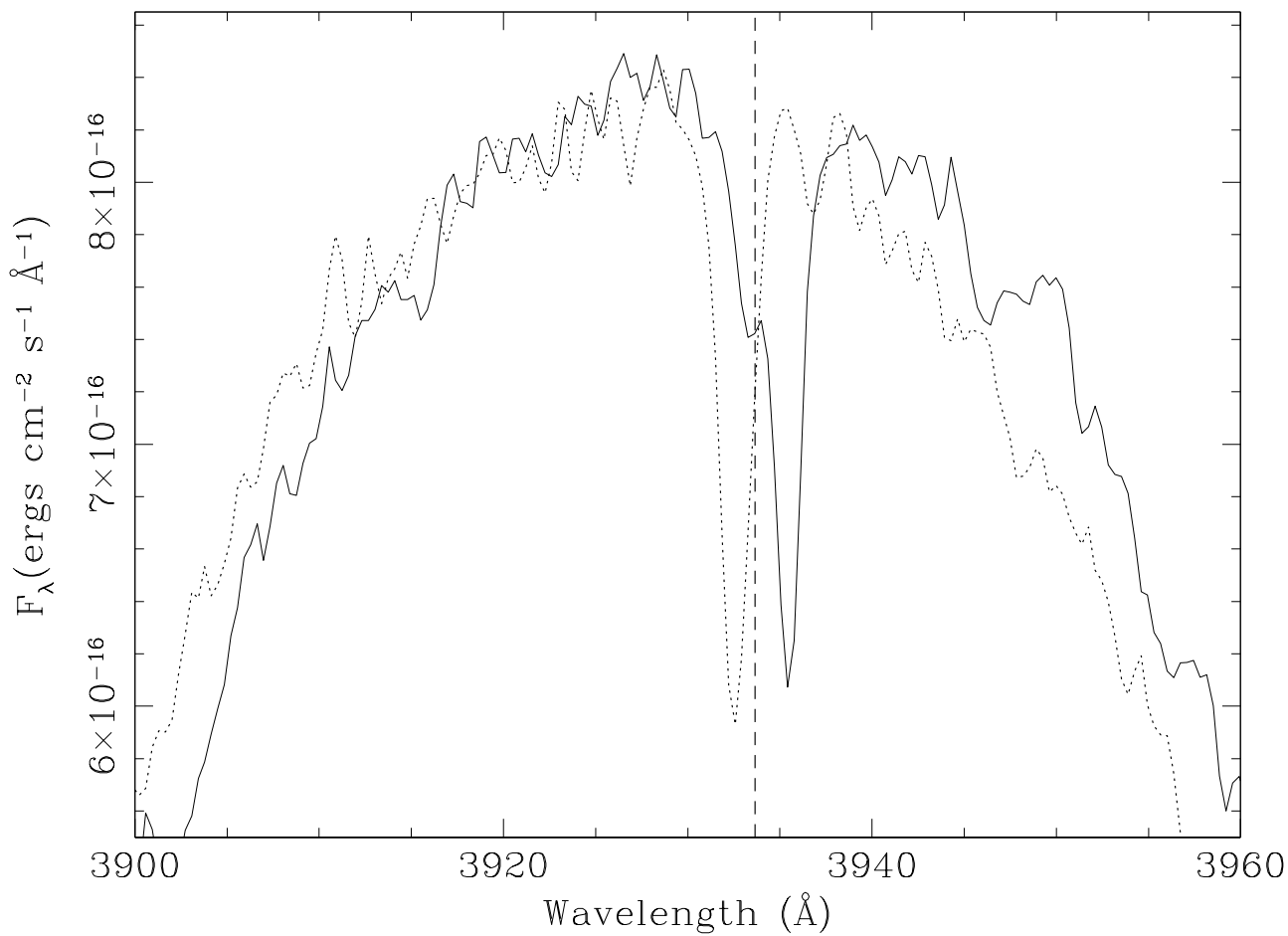} 
\caption{The spectra averaged around the maximum and minimum radial velocity for J0917+46. The red-shifted spectrum (solid line) is a combination of five spectra with $V =$ 125,
134, 152, 160, and 173 km s$^{-1}$ shifted to an average velocity of 149 km s$^{-1}$. The blue-shifted spectrum (dotted line) is a combination of four spectra with
$V= -80, -83, -87$, and $-90$ km s$^{-1}$ shifted to an average velocity of $-$85 km s$^{-1}$. The dashed line marks the rest wavelength of the Ca K line.}
\end{figure}

\end{document}